\documentclass[12pt]{article}
\usepackage{psfig}
\def\graphpath{}

\begin{document}
\begin{flushright}
OHSTPY-HEP-T-99-024 \\
UMD-D-99-4 \\
hep-th/9911243
\end{flushright}
\vspace{20mm}
\begin{center}
%{\LARGE The Solution of N=1 SYM in (2+1) Dimensions}
{\LARGE The Mass Spectrum of ${\cal N}=1$ SYM$_{2+1}$ at Strong Coupling}
\\
\vspace{20mm}
{\bf Paul Haney$^a$, John R. Hiller$^b$, Oleg Lunin$^a$,\\ Stephen
Pinsky$^a$, Uwe
Trittmann$^a$}
\\
\vspace{4mm}
{\em $^a$Department of Physics,\\ The Ohio State University,\\ Columbus,
OH 43210, USA\\ \vspace{4mm}
$^b$Department of Physics,\\ University of Minnesota Duluth,\\ Duluth, MN
55812, USA\\}

\end{center}
\vspace{10mm}

\begin{abstract}
We consider supersymmetric Yang--Mills theory on
${\bf R} \times S^1 \times S^1$. In particular, we choose one of the compact
directions to be light--like and another to be space--like. Since the SDLCQ
regularization explicitly preserves supersymmetry, this theory is
totally finite, and
thus we can solve for bound state wave
functions and masses numerically without renormalizing.  We present the masses
as functions of the longitudinal and transverse resolutions and show that the
masses converge rapidly in both resolutions. We also study the behavior of the
spectrum as a function of the coupling and find that at strong
coupling there is
a stable, well defined spectrum which we present. We also find several
unphysical
states that decouple at large transverse resolution. There are two sets of
massless states; one set is massless only at zero coupling and the other is
massless at all couplings.  Together these sets of massless states are in
one--to--one correspondence with the full spectrum of the
dimensionally reduced
theory.

\end{abstract}
\newpage

\def\be{\begin{equation}}
\def\bea{\begin{eqnarray}}
\def\ee{\end{equation}}
\def\eea{\end{eqnarray}}
\def\d{\partial}

\baselineskip .25in

%%%%%%%%%%%%%%%%%%%%%%%%%%%%%%%%%%%%%%%%%%%%%%
\section{Introduction}
Recently, there has been considerable
progress in understanding the properties of strongly coupled
gauge theories with supersymmetry
\cite{seibergwitten,seiberg,maldacena}. In particular, there
are a number of supersymmetric gauge theories that are believed
to be interconnected through a web of strong-weak coupling dualities.
Although existing evidence for
these dualities is encouraging, there is still an urgent need
to address these issues at a more fundamental level. Ideally,
we would like to solve for the bound states of these
theories directly, and at any coupling.

Of course, solving a field theory from first principles
is typically an intractable task. Nevertheless, it has been
known for some time that $1+1$ dimensional field theories
{\em can} be solved from first principles via a straightforward
application of
DLCQ  (see \cite{bpp98} for a review).
In more recent times, a large class of supersymmetric
gauge theories in two dimensions was studied using
a supersymmetric form of DLCQ  (or `SDLCQ'), which is known to
preserve supersymmetry \cite{sakai95,klebhash,alp98a,alp98b,
alpp98,alpp99,alp99}.

We have recently been able to extend
the SDLCQ algorithms to solve higher-dimensional theories \cite{alp99b}.
One important difference between two-dimensional and
higher-dimensional theories is the phase diagram induced by
variations in the gauge coupling. The
spectrum of a $1+1$ dimensional gauge theory
scales trivially with
respect to the gauge coupling, while a theory in higher dimensions
has the potential of exhibiting a complex phase structure,
which may include a strong-weak coupling duality.
Ref.~\cite{alp99b} seemed to provide hints of the latter.
It is therefore interesting to study
the phase diagram of gauge theories in $D \geq 3$ dimensions.

Towards this end, we consider  three dimensional SU($N_c$) ${\cal N}=1$
super-Yang-Mills   compactified on the space-time ${\bf R} \times S^1 \times
S^1$. The calculations are all done in the large $N_c$ limit.  In
particular, we
compactify the light-cone coordinate
$x^-$ on a light-like circle via DLCQ, and wrap the remaining  transverse
coordinate $x^{\perp}$ on a spatial circle.   We are able to solve
for bound state
wave functions and masses numerically by diagonalizing the  discretized
light-cone supercharge. We have shown that the SDLCQ procedure   extends
naturally to $2+1$ dimensions, resulting in an exactly supersymmetric
spectrum.

The contents of this paper are organized as follows.
In Section \ref{formulation}, we formulate
SU($N_c$) ${\cal N}=1$ super-Yang-Mills
defined on the compactified space-time ${\bf R} \times S^1 \times S^1$.
Explicit expressions are given for the light-cone supercharges,
which are then discretized via the SDLCQ procedure.
Quantization of the theory is carried out by imposing
canonical \mbox{(anti-)}commutation relations for boson and fermion
fields. We also
discuss the two discrete symmetries of the theory in this section. In Section
\ref{numerical}, we present our numerical results. We present plots of the
spectrum as  a function of the longitudinal and transverse  resolution and
show that it convergences very rapidly in both.
We plot the spectrum also as a
function of the coupling and find a very stable strong-coupling spectrum. We
identify several states that appear to be unphysical and disappear at high
transverse resolutions. Finally we discuss the infrared spectrum of the theory
and present evidence that it is entirely determined by the dimensionally
reduced theory. We conclude our analysis with a discussion of our results and a
list of important related future projects in Section \ref{summary}.
 
%%%%%%%%%%%%%%%%%%%%%%%%%%%%%%%%%%%%%%%%%%%%%%%
\section{Light-Cone Quantization and SDLCQ}
\label{formulation}

Discrete light cone quantization has proven to be a very powerful method of
studying the mass spectra of various theories \cite{bpp98}.
It is well known that in $1+1$ dimension one can define a new version of DLCQ
which preserves supersymmetry \cite{sakai95,alp99}. In higher dimensions a
supersymmetric prescription is also possible \cite{alp99b}. We begin by
introducing light-cone coordinates
$x^{\pm} = (x^0 \pm x^1)/\sqrt{2}$, and compactifying the $x^-$ coordinate on a
light-like circle. In this way, the conjugate light-cone momentum
$k^+$ is discretized.
To discretize the remaining (transverse) momentum $k^{\perp} = k^2$, we may
compactify $x^{\perp} = x^2$ on a spatial circle.  Of course, there
is a significant
difference between the discretized light-cone momenta $k^+$ and the discretized
transverse momenta
$k_{\perp}$; namely, the light-cone momentum $k^+$ is always positive, while
$k_{\perp}$ may take on positive or negative values. The positivity
of $k^+$ is a key
property that is exploited in DLCQ calculations;  for any given light-cone
compactification, there are only a finite number of choices for $k^+$
-- the total
number depending on how finely we discretize the  momenta. The `resolution'
of the discretization is usually characterized by a positive integer
$K$, which is called
the `harmonic resolution'
\cite{pb85,yamawaki}; for a given choice of $K$, the light-cone
momenta $k^+$ are
restricted to positive integer multiples of $P^+/K$, where $P^+$ is
the total light-cone
momentum of a state. In the context of two-dimensional theories,
this implies a finite
number of Fock states \cite{pb85}.

In the case of interest here, we include   %we are interested in there is
an additional transverse
dimension, and the number of Fock states is no longer finite,
since there is an arbitrarily large number of transverse momentum
modes defined on the transverse spatial circle.
Thus, an additional truncation of the transverse momentum
modes is required to render the total number of Fock states
finite and the problem numerically tractable.
This truncation procedure, which is characterized by a transverse
resolution T, is
analogous to the truncation of $k^+$  imposed by the `harmonic resolution' $K$.
Thus the transverse momentum $k_\perp$ range from zero to $\pm\frac{2
\pi T}{L}$ ,
where $L$ is the size of the transverse circle.

Let us now review these ideas in the context of a specific
super-Yang-Mills theory.
We start with $(2+1)$-dimensional ${\cal N}=1$ super-Yang-Mills theory
defined on a space-time with one transverse dimension
compactified on a circle:
\be
S=\int d^2 x \int_0^L dx_\perp \mbox{tr}(-\frac{1}{4}F^{\mu\nu}F_{\mu\nu}+
{\rm i}{\bar\Psi}\gamma^\mu D_\mu\Psi).
\ee
After introducing the light--cone coordinates
$x^\pm=\frac{1}{\sqrt{2}}(x^0\pm x^1)$, decomposing the spinor $\Psi$
in terms of chiral projections
\be
\psi=\frac{1+\gamma^5}{2^{1/4}}\Psi,\qquad
\chi=\frac{1-\gamma^5}{2^{1/4}}\Psi,
\ee
and choosing the light--cone gauge $A^+=0$, we obtain   %the action becomes
\bea\label{action}
S&=&\int dx^+dx^- \int_0^L dx_\perp \mbox{tr}\left[\frac{1}{2}(\d_-A^-)^2+
(D_+\phi+\d_\perp A^-)\d_-\phi+ {\rm i}\psi D_+\psi+ \right.\nonumber \\
& &
\left.
         \hspace{15mm} +{\rm i}\chi\d_-\chi+\frac{{\rm i}}{\sqrt{2}}\psi
D_\perp\chi+
\frac{{\rm i}}{\sqrt{2}}\chi D_\perp\psi \right].
\eea
A simplification of the
light--cone gauge is that the
non-dynamical fields $A^-$ and $\chi$ may be explicitly
solved from their Euler-Lagrange equations of motion:
\be
A^-=\frac{1}{\d_-^2}J=
\frac{1}{\d_-^2}\left(ig[\phi,\d_-\phi]+2g\psi\psi -\d_\perp \d_-\phi
\right), \quad
\chi=-\frac{1}{\sqrt{2}\d_-}D_\perp\psi.\nonumber
\ee
These expressions may be used to express any operator
in terms of the physical degrees of freedom only.
In particular, the light-cone energy, $P^-$, and momentum
operators, $P^+$,$P^{\perp}$,
corresponding to  translation
invariance in each of the coordinates
$x^\pm$ and $x_\perp$ may be calculated explicitly:
\bea\label{moment}
P^+&=&\int dx^-\int_0^L dx_\perp\mbox{tr}\left[(\d_-\phi)^2+
{\rm i}\psi\d_-\psi\right],\\
P^-&=&\int dx^-\int_0^L dx_\perp\mbox{tr}
\left[-\frac{1}{2}J\frac{1}{\d_-^2}J-
            \frac{{\rm i}}{2}D_\perp\psi\frac{1}{\d_-}D_\perp\psi\right],\\
P_\perp &=&\int dx^-\int_0^L dx_\perp\mbox{tr}\left[\d_-\phi\d_\perp\phi+
            {\rm i}\psi\d_\perp\psi\right].
\eea
The light-cone supercharge in this theory
is a two component Majorana spinor, and may be conveniently
decomposed in terms of its chiral projections:
\bea\label{sucharge}
Q^+&=&2^{1/4}\int dx^-\int_0^L dx_\perp\mbox{tr}\left[\phi\d_-\psi-\psi\d_-
                   \phi\right],\\
\label{sucharge-}
Q^-&=&2^{3/4}\int dx^-\int_0^L dx_\perp\mbox{tr}\left[\d_\perp\phi\psi+
            g\left({\rm
i}[\phi,\d_-\phi]+2\psi\psi\right)\frac{1}{\d_-}\psi\right].
\eea
The action (\ref{action}) gives the following canonical
(anti-)commutation relations for
propagating fields for large $N_c$ at equal $x^+$:
\begin{equation}
\left[\phi_{ij}(x^-,x_\perp),\d_-\phi_{kl}(y^-,y_\perp)\right]=
\left\{\psi_{ij}(x^-,x_\perp),\psi_{kl}(y^-,y_\perp)\right\}=
\frac{1}{2}\delta(x^- -y^-)\delta(x_\perp -y_\perp)\delta_{il}\delta_{jk}.
\label{comm}
\end{equation}
Using these relations one can check the supersymmetry algebra
\be
\{Q^+,Q^+\}=2\sqrt{2}P^+,\qquad \{Q^-,Q^-\}=2\sqrt{2}P^-,\qquad
\{Q^+,Q^-\}=-4P_\perp.
\label{superr}
\ee

We will consider only states which have vanishing transverse momentum,
which is possible since the total transverse momentum operator
is kinematical\footnote{Strictly speaking, on a transverse
cylinder, there are separate sectors with total
transverse momenta $2\pi n/L$; we consider only one of them, $n=0$.}.
On such states, the light-cone supercharges
$Q^+$ and $Q^-$ anticommute with each other, and the supersymmetry algebra
is equivalent to the ${\cal N}=(1,1)$ supersymmetry
of the dimensionally reduced ({\em i.e}.\ two-dimensional) theory
\cite{sakai95}.
Moreover, in the $P_{\perp} = 0$ sector,
the mass squared operator $M^2$ is given by
$M^2=2P^+P^-$.

As we mentioned earlier, in order to render the bound state equations
numerically tractable, the transverse
momentum of partons must be truncated.
First, we introduce the Fourier expansion for the fields $\phi$ and $\psi$,
where the transverse space-time coordinate $x^{\perp}$ is periodically
identified:
\bea
\lefteqn{
\phi_{ij}(0,x^-,x_\perp) =} & & \nonumber \\
& &
\frac{1}{\sqrt{2\pi L}}\sum_{n^{\perp} = -\infty}^{\infty}
\int_0^\infty
         \frac{dk^+}{\sqrt{2k^+}}\left[
         a_{ij}(k^+,n^{\perp})e^{-{\rm i}k^+x^- +{\rm i}
\frac{2 \pi n^{\perp}}{L} x_\perp}+
         a^\dagger_{ji}(k^+,n^{\perp})e^{{\rm i}k^+x^- -
{\rm i}\frac{2 \pi n^{\perp}}{L}  x_\perp}\right]
\nonumber\\
\lefteqn{
\psi_{ij}(0,x^-,x_\perp) =} & & \nonumber \\
& & \frac{1}{2\sqrt{\pi L}}\sum_{n^{\perp}=-\infty}^{\infty}\int_0^\infty
         dk^+\left[b_{ij}(k^+,n^{\perp})e^{-{\rm i}k^+x^- +
{\rm i}\frac{2 \pi n^{\perp}}{L} x_\perp}+
         b^\dagger_{ji}(k^+,n^\perp)e^{{\rm i}k^+x^- -{\rm i}
\frac{2 \pi n^{\perp}}{L} x_\perp}\right]
\nonumber
\eea
Substituting these into the (anti-)commutators (\ref{comm}),
one finds
\begin{equation}
\left[a_{ij}(p^+,n_\perp),a^\dagger_{lk}(q^+,m_\perp)\right]=
\left\{b_{ij}(p^+,n_\perp),b^\dagger_{lk}(q^+,m_\perp)\right\}=
\delta(p^+ -q^+)\delta_{n_\perp,m_\perp}\delta_{il}\delta_{jk}.
\end{equation}
The supercharges now take the following forms:
\bea\label{TruncSch}
&&Q^+={\rm i}2^{1/4}\sum_{n^{\perp}\in {\bf Z}}\int_0^\infty dk\sqrt{k}\left[
b_{ij}^\dagger(k,n^\perp) a_{ij}(k,n^\perp)-
a_{ij}^\dagger(k,n^\perp) b_{ij}(k,n^\perp)\right],\\
\label{Qminus}
&&Q^-=\frac{2^{3/4}\pi {\rm i}}{L}\sum_{n^{\perp}\in {\bf Z}}\int_0^\infty dk
\frac{n^{\perp}}{\sqrt{k}}\left[
a_{ij}^\dagger(k,n^\perp) b_{ij}(k,n^\perp)-
b_{ij}^\dagger(k,n^\perp) a_{ij}(k,n^\perp)\right]+\nonumber\\
&&+ {{\rm i} 2^{-1/4} {g} \over \sqrt{L\pi}}
\sum_{n^{\perp}_{i} \in {\bf Z}} \int_0^\infty dk_1dk_2dk_3
\delta(k_1+k_2-k_3) \delta_{n^\perp_1+n^\perp_2,n^\perp_3}\nonumber\\
&&\times\left\{
     {1 \over 2\sqrt{k_1 k_2}} {k_2-k_1 \over k_3}
[a_{ik}^\dagger(k_1,n^\perp_1) a_{kj}^\dagger(k_2,n^\perp_2)
b_{ij}(k_3,n^\perp_3)
-b_{ij}^\dagger(k_3,n^\perp_3)a_{ik}(k_1,n^\perp_1)
a_{kj}(k_2,n^\perp_2) ]\right.\nonumber\\
&&+{1 \over 2\sqrt{k_1 k_3}} {k_1+k_3 \over k_2}
[a_{ik}^\dagger(k_3,n^\perp_3) a_{kj}(k_1,n^\perp_1) b_{ij}(k_2,n^\perp_2)
-a_{ik}^\dagger(k_1,n^\perp_1) b_{kj}^\dagger(k_2,n^\perp_2)
a_{ij}(k_3,n^\perp_3) ]\nonumber\\
&&+{1 \over 2\sqrt{k_2 k_3}} {k_2+k_3 \over k_1}
[b_{ik}^\dagger(k_1,n^\perp_1) a_{kj}^\dagger(k_2,n^\perp_2)
a_{ij}(k_3,n^\perp_3)
-a_{ij}^\dagger(k_3,n^\perp_3)b_{ik}(k_1) a_{kj}(k_2,n^\perp_2) ]\nonumber\\
&&+({ 1\over k_1}+{1 \over k_2}-{1\over k_3})
[b_{ik}^\dagger(k_1,n^\perp_1) b_{kj}^\dagger(k_2,n^\perp_2)
b_{ij}(k_3,n^\perp_3)
+b_{ij}^\dagger(k_3,n^\perp_3) b_{ik}(k_1,n^\perp_1) b_{kj}(k_2,n^\perp_2)]
         \left. \frac{}{}\right\}. \nonumber \\
\eea
We now perform the truncation procedure; namely, in all sums over the
transverse momenta $n^{\perp}_{i}$ appearing in the above expressions for the
supercharges, we restrict summation to the following allowed momentum
modes: $n^{\perp}_{i}=0,\pm 1 ... \pm T$.  Note that this prescription is
symmetric, in the sense that there are as many positive modes as there are
negative ones. In this way we  retain parity symmetry in the transverse
direction.

How does such a truncation affect the supersymmetry properties of the
theory? Note first that an operator relation $[A,B]=C$ in the  full
theory is not
expected to hold in the truncated formulation.  However, if A is quadratic in
terms of fields (or in terms of creation and  annihilation operators), one can
show that the relation $[A,B]=C$ implies
$$
[A_{tr},B_{tr}]=C_{tr}
$$
for the truncated operators $A_{tr}$,$B_{tr}$, and $C_{tr}$.  In our
case, $Q^+$
is quadratic, and so the relations
$\{Q_{tr}^+,Q_{tr}^+\}=2\sqrt{2}P_{tr}^+$ and
$\{Q_{tr}^+,Q_{tr}^-\}=0$ are true
in the $P_\perp=0$ sector of the truncated theory.  The anticommutator
$\{Q_{tr}^-,Q_{tr}^-\}$,
however, is not equal to $2\sqrt{2}P_{tr}^-$. So the diagonalization of
$\{Q_{tr}^-,Q_{tr}^-\}$ will yield a different bound-state spectrum
than the one
obtained after diagonalizing $2\sqrt{2}P_{tr}^-$. Of course, the two spectra
should agree in the limit
$T\rightarrow\infty$. At any finite truncation, however, we have the liberty to
diagonalize either of these operators. The choice of $\{Q_{tr}^-,Q_{tr}^-\}$
specifies our regularization scheme.

Choosing to diagonalize the light-cone
supercharge $Q_{tr}^-$ has an important advantage:
{\em the spectrum is exactly supersymmetric for
any truncation}. In contrast, the spectrum of the Hamiltonian $P_{tr}^-$
becomes supersymmetric only in the infinite resolution limit.

Let us discuss the discrete symmetries of $Q^-$. There are three commuting
$Z_2$ symmetries, one of them is the parity in the transverse direction:
\begin{equation}\label{defparity}
P: a_{ij}(k,n^\perp)\rightarrow -a_{ij}(k,-n^\perp),\qquad
        b_{ij}(k,n^\perp)\rightarrow b_{ij}(k,-n^\perp).
\end{equation}
Note that $P$ does not commute with $Q^+$ or with $P_\perp$:
\begin{equation}
PQ^+=-Q^+P, \qquad PP_\perp=-P_\perp P.
\end{equation}
The second symmetry is given by a generalization of the T--symmetry defined in
\cite{kutasov93} (we will call it S to avoid the confusion with transverse
truncation parameter):
\begin{equation}
\label{defz2}
S: a_{ij}(k,n^\perp)\rightarrow -a_{ji}(k,n^\perp),\qquad
        b_{ij}(k,n^\perp)\rightarrow -b_{ji}(k,-n^\perp).
\end{equation}
It commutes with all the Noether charges (\ref{moment})--(\ref{sucharge-}).
Since $P$ and $S$ commute with each other one needs only one additional
symmetry $R=PS$ to close the group.

Since $Q^-$, $P$ and $S$ commute with each other we can diagonalize them
simultaneously. This allows us to diagonalization of the supercharge
separately
in the sectors with fixed $P$ and $S$ parities and thus will reduce the size
of matrices. Doing this one finds that the roles of $P$ and $S$ are different.
While all the eigenvalues are usually broken into non-overlapping $S$--odd and
$S$--even sectors \cite{bdk93}, the $P$ symmetry leads to a double
degeneracy of massive states (in addition to usual boson--fermion degeneracy
due to supersymmetry). To demonstrate this, let us start from the massive
bosonic state $|M+\rangle$:
\begin{equation}
\left(Q^-\right)^2|M+\rangle=M^2|M+\rangle,\qquad
P|M+\rangle=+|M+\rangle.
\end{equation}
If $M\ne 0$ then the bosonic state $Q^+Q^-|M+\rangle$ has nonzero norm and it
is also an eigenstate of $\left(Q^-\right)^2$ with eigenvalue $M^2$. This state
however has a negative parity:
\begin{equation}
PQ^+Q^-|M+\rangle=-Q^+PQ^-|M+\rangle=-Q^+Q^-P|M+\rangle=-Q^+Q^-|M+\rangle.
\end{equation}
Thus we found one more nice feature of SDLCQ: in addition to preserving
supersymmetry in the truncated theory it also preserves the degeneracy of the
spectrum related to parity. We believe that this degeneracy will be lost
in the usual DLCQ approximation but currently have no proof of that.

To summarize, we have introduced a truncation procedure
that facilitates a numerical study of the bound state problem, and
preserves supersymmetry.
The interesting property of the light-cone supercharge $Q^-$
(\ref{Qminus}) is the
presence of a gauge coupling constant as an independent variable,
which does not appear in the study of two-dimensional theories.
In the next section, we will study how
variations in this coupling affects the bound states
in the theory.

%%%%%%%%%%%%%%%%%%%%%%%%%%%%%%%%%%%%%%%%%%%%%%%%%%%
\section{Numerical Results}
\label{numerical}
In order to implement the DLCQ formulation of the bound-state problem -- which
is tantamount to imposing periodic boundary conditions
$x^- = x^- + 2 \pi R$ \cite{yamawaki} -- we simply restrict the
light-cone momentum variables $k_i$ appearing in the expressions for $Q^+$
and $Q^-$ to the following discretized set of momenta:
$\left\{ \frac{1}{K} P^+, \frac{2}{K} P^+, \frac{3}{K} P^+,\dots,
\right\}$. Here, $P^+$ denotes the total light-cone momentum of a
state, and may be thought of as a fixed constant, since it is
easy to form a Fock basis that is already diagonal with respect to the
operator $P^+$ \cite{pb85}.
%
%The integer $K$ is called the `harmonic resolution', and $1/K$
%
The reciprocal of the harmonic resolution $K$
measures the coarseness of our discretization.
The continuum limit is then recovered by taking the limit
$K \rightarrow \infty$. Physically, $1/K$ represents the smallest
positive unit of longitudinal momentum fraction allowed for each
parton in a Fock state.

%%%%%%%%%%%%%%%%%%%%%%%%%%
\begin{figure}[ht]
\centerline{
\psfig{file=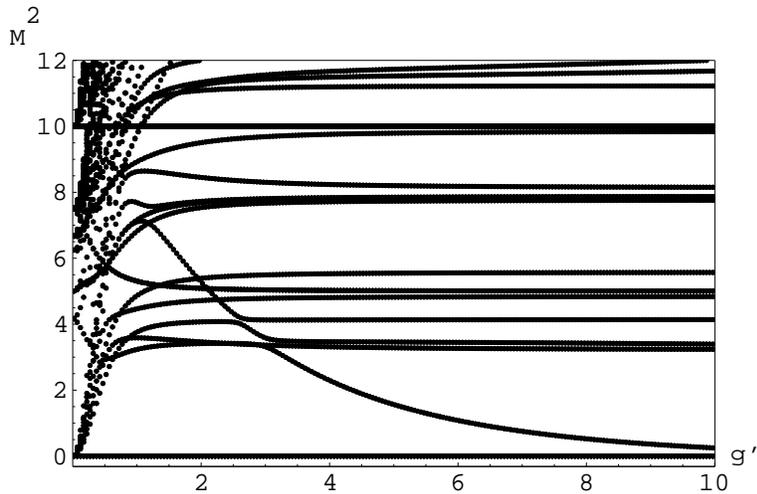,width=10.0true cm,angle=0}
}
\caption{{\small Plot of bound state mass squared $M^2$
in units of $2 \pi^2 / L^2$
as a function of the dimensionless coupling $0 \leq g' \leq 10$,
defined by $(g')^2 = g^2 N L/4 \pi^3$,
at $N=1000$ and $K=5$. Note that there are massless states.} \label{old}}
\end{figure}
%%%%%%%%%%%%%%%%%%%%%%%%%%

Of course, as soon as we implement the SDLCQ procedure, which is  specified
unambiguously by the harmonic resolution $K$,  and cut off transverse  momentum
modes via the constraint $|n_i^{\perp}| \leq T$,  the integrals  appearing in
the definitions  for $Q^+$ and $Q^-$ are replaced by finite sums, and so  the
eigen-equation
$2P^+P^-|\Psi\rangle = M^2 |\Psi\rangle$ is reduced to a finite  matrix
diagonalization problem. In this last step we use the fact that $P^-$ is
proportional  to the square of the light-cone supercharge\footnote{ Strictly
speaking, $P^- =
\frac{1}{\sqrt{2}}(Q^-)^2$ is an identity in the continuum theory, and a {\em
definition} in the compactified theory, corresponding to the SDLCQ prescription
\cite{sakai95,alp99}.} $Q^-$. Previously \cite{alp99b} we studied
this theory with
$K$ up to $5$, but with only one unit of transverse momentum corresponding to
$T=1$.  In Figure \ref{old} we show the spectrum we obtained in
that study as a function of a dimensionless coupling
$g'=g\sqrt{NL/4\pi^3}$.  This figure shows several striking
features that we want to analyze in more detail. First, we want to know
whether the well defined strong-coupling spectrum, observed in
Figure \ref{old},
persists at higher values of transverse resolution $T$ and study its
convergence in $T$. The other striking feature of the $K=5$, $T=1$ spectrum is
the state that is falling rapidly at large coupling. We will analyze the
behavior of analogous states at different values of longitudinal and
transverse resolutions and in particular we will be interested in the fate of
the asymptotic (as $g\rightarrow\infty$) massless state in the continuum
limit.
In addition to this state there are many states that stay exactly massless
at all values of coupling. As we have shown in \cite{alp99b}, the number of
such states does not depend on the value of the transverse resolution.  The new
numerical data support this result. The massive state at $g=0$ are just the
discrete approximation the a continuum of free particles and is extactly
calculable analytically.

Our previous SDLCQ calculations were done using a code written in
Mathematica and performed on a PC. This code has now been rewritten in
C++ and some of the present work was done on supercomputers. We were able
to perform numerical diagonalizations for $K=2$ through 7 and for values of
$T$ up to $T=9$ at
$K=4$ and $T=1$ at $K=7$.

%%%%%%%%%%%%%%%%%%%%%%%%%%
\begin{figure}[ht]
\centerline{
\psfig{file=\graphpath 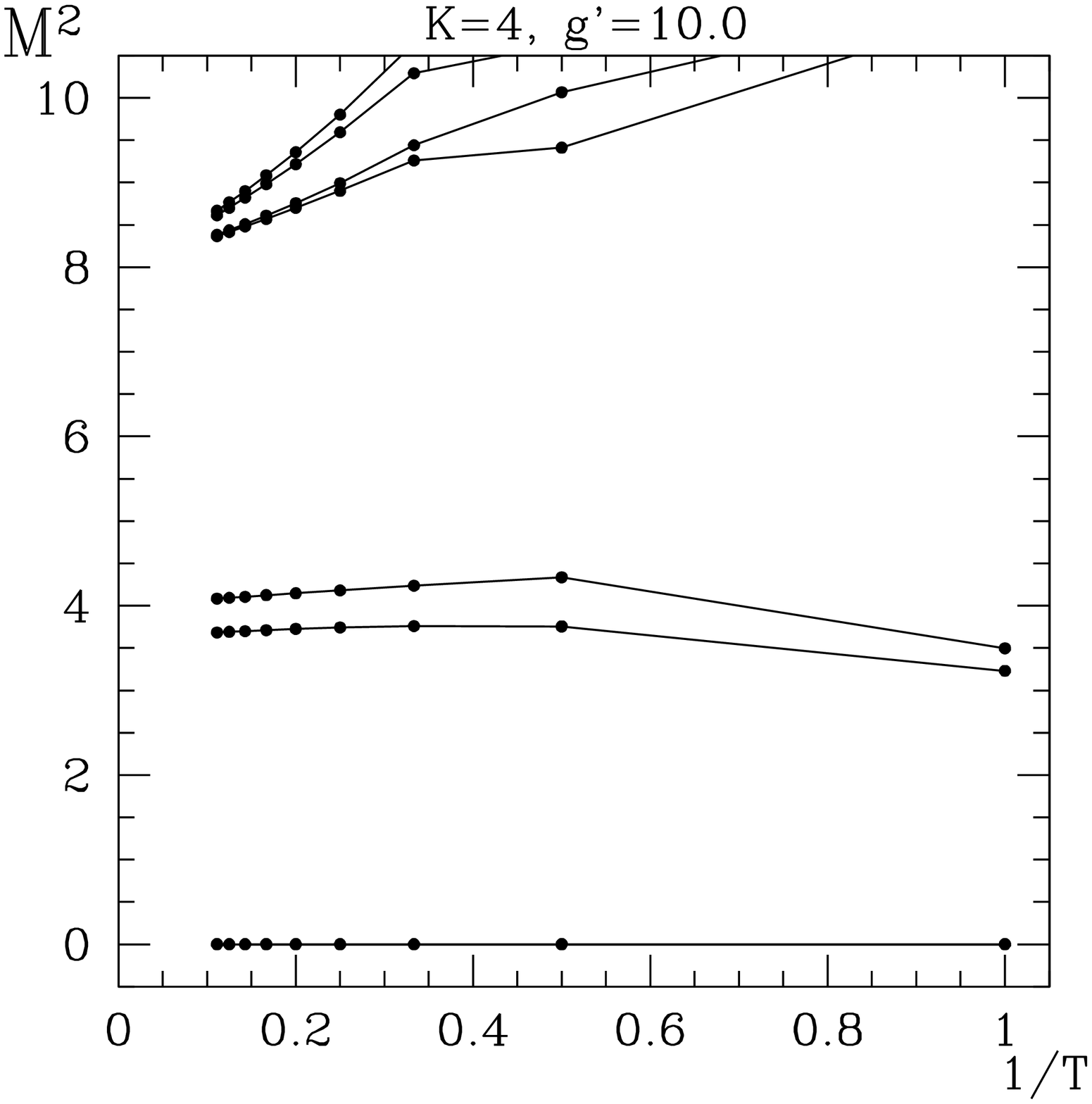,width=8.0true cm,angle=0}
\psfig{file=\graphpath 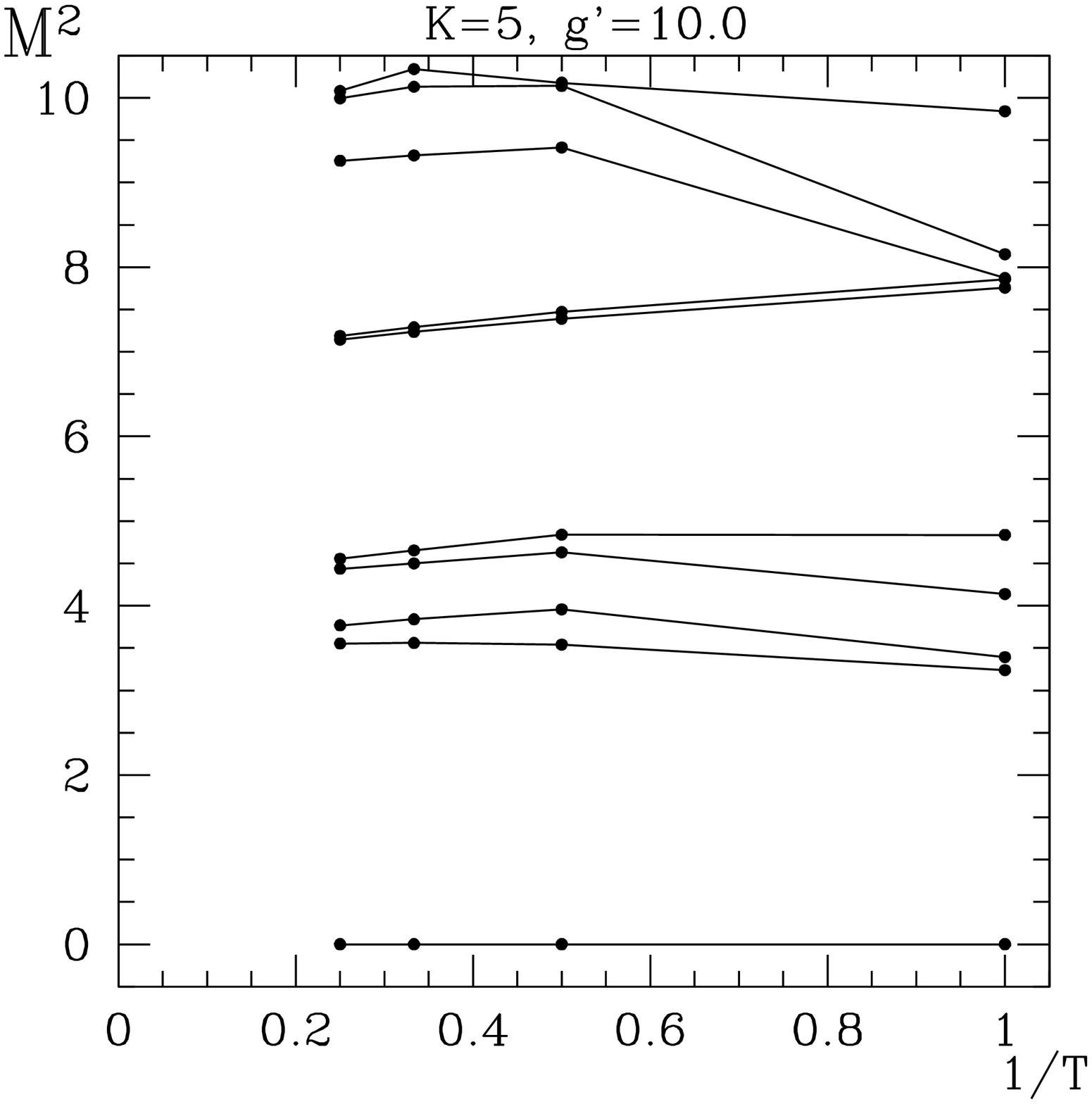,width=8.0true cm,angle=0}
}
\caption{{\small Plot of bound state mass squared $M^2$
in units of $2 \pi^2/ L^2$  as a function of the transverse
resolution $T$ for a coupling
$g'=10$ and for longitudinal resolutions $K=4$ (a) and $K=5$ (b).
Boson and fermion masses are identical. }
\label{perpmom}}
\end{figure}
%%%%%%%%%%%%%%%%%%%%%%%%%%

%%%%%%%%%%%%%%%%%%%%%%%%%%
\begin{figure}[ht]
\centerline{
\psfig{file=\graphpath 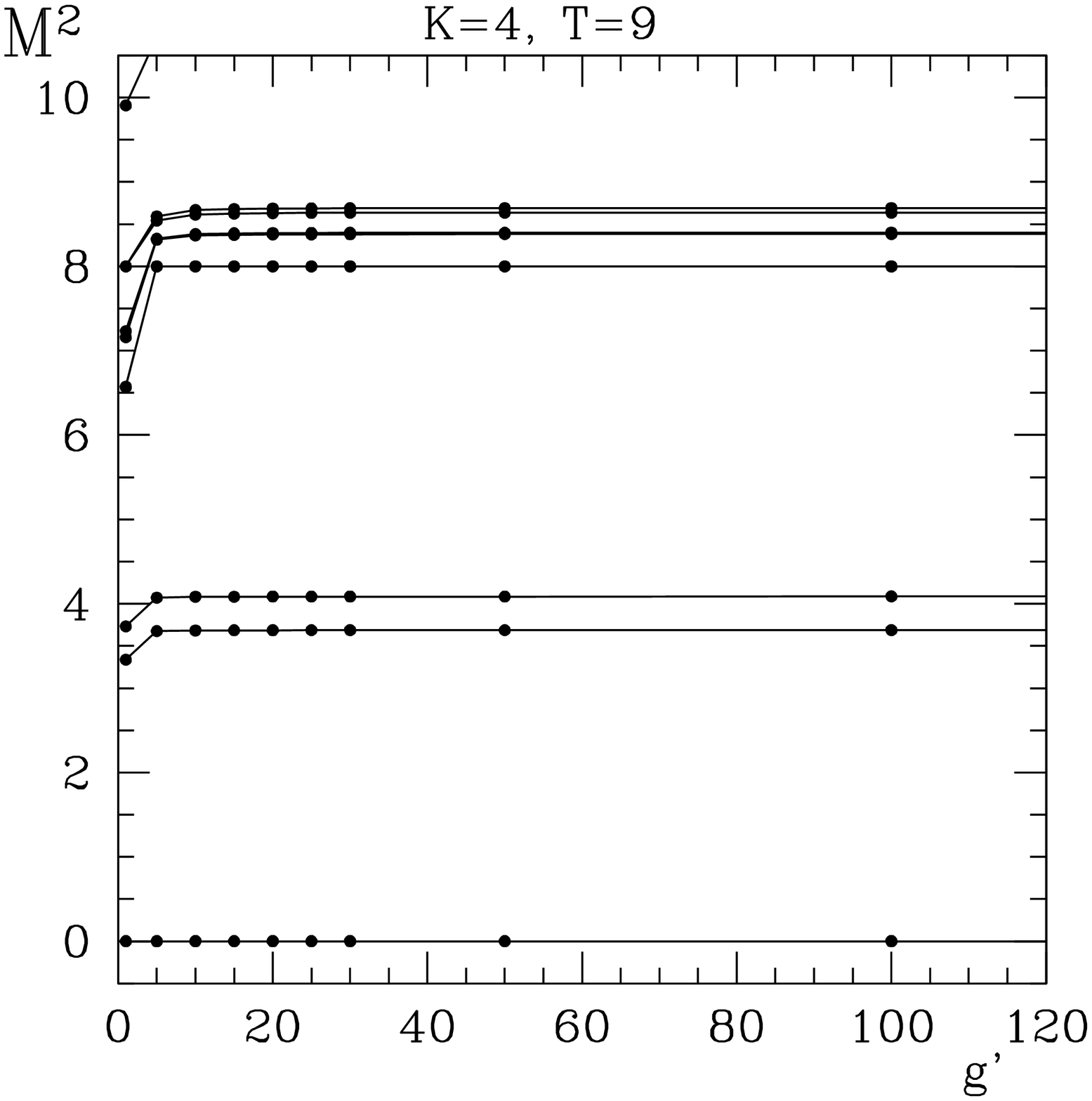,width=8.0true cm,angle=0}
\psfig{file=\graphpath 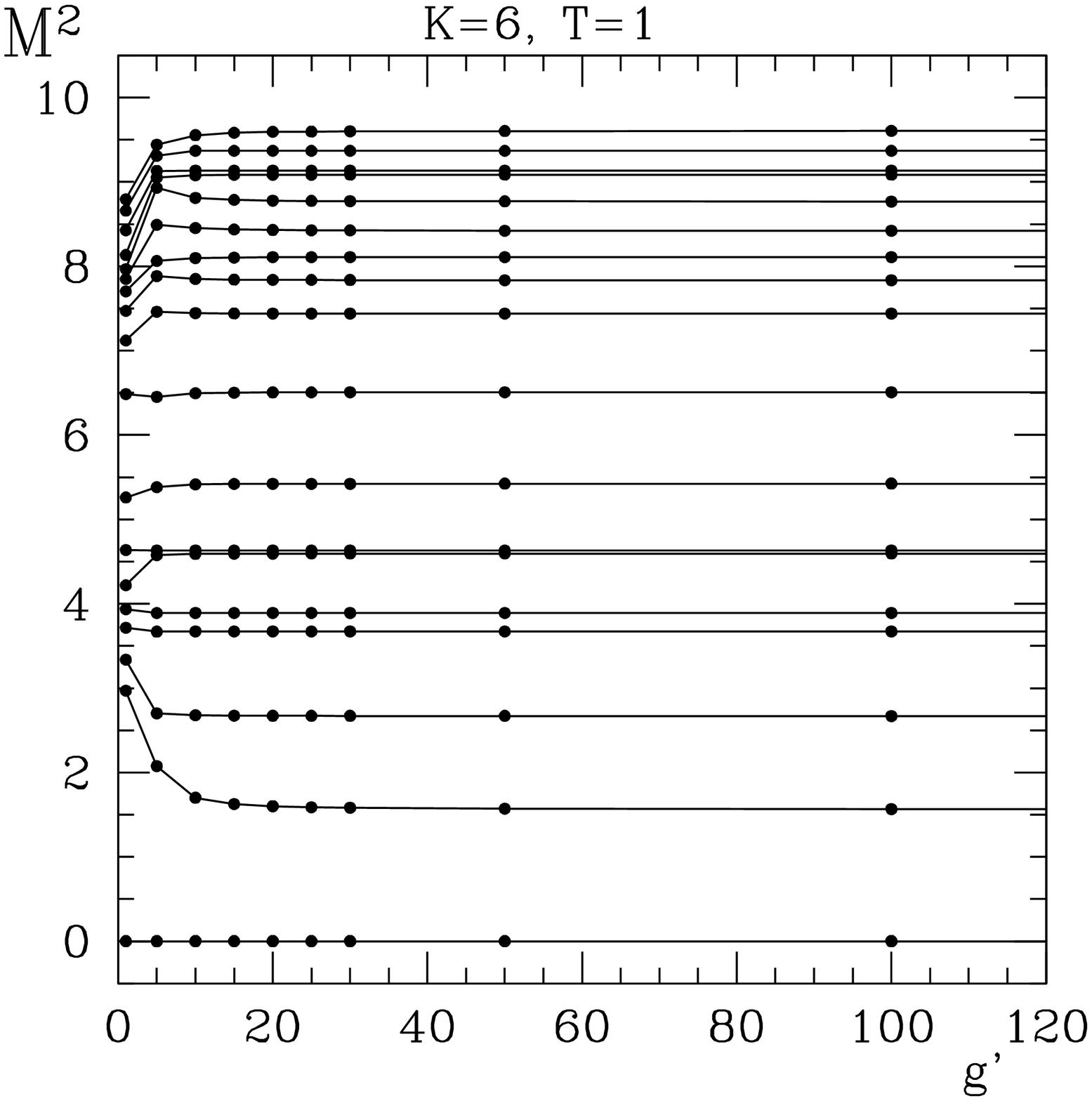,width=8.0true cm,angle=0}
}
\caption{{\small  Plot of bound state mass squared $M^2$
in units of $4 \pi^2 / L^2$  as a function of the coupling
$g'$.  We show the plots for $K=4$, $T=9$ (a) and $K=6$,
$T=1$ (b).}
\label{coupling}}
\end{figure}
%%%%%%%%%%%%%%%%%%%%%%%%%%

%%%%%%%%%%%%%%%%%%%%%%%%%%%%%%%%%%%%%%%%%%%%%%%%%%%%%%%%%%%%%
\subsection{Strong Coupling: Massive Spectrum}
There are very few theories, other than in $1+1$ dimensions, where we have good
information about the  strong-coupling
spectrum. In $1+1$ dimensions the concept of strong coupling has
very little meaning since the coupling only comes in as an overall
multiplicative
constant in the Hamiltonian. There are a few lattice results but
most of what we
know comes from theories that have dualities with weakly coupled theories. For
the $(2+1)$-dimensional ${\cal N}=1$ SYM
theory we are considering, there is no  known duality between strong
and weak coupling. Using SDLCQ, however, we can directly uncover the
behavior of
the theory at strong coupling.

In Figure \ref{perpmom}, we plot the bound state mass squared
$M^2$, in units of $4 \pi^2 /L^2$, as a function of the transverse
resolution $T$\ for $K=4$ and $K=5$ in the strong-coupling
regime. We see that
these curves are amazingly flat, showing that this theory converges
very rapidly with
the transverse cutoff.  We saw previously in $(1+1)$-dimensional
models \cite{alp98a} that SDLCQ  converges much
faster than DLCQ, and this
appears to persist for the transverse momentum
in $2+1$ dimensions. From these figures it
appears that we get sensible results by $T=2$ and good results already
for  $T=3$.

Given this rapid convergence, a sensible procedure is to remove the
transverse cutoff by extrapolating the masses of the low lying
bounds states to
large transverse resolution at each value of the longitudinal resolution
(thus constructing the SDLCQ spectrum of the complete three dimensional theory)
and then extrapolating to large longitudinal resolution K for each of the
states to find the spectrum as a function of the coupling.  For $K=6$ and $7$
we simply take the masses at the largest transverse momentum since there are
not enough transverse moment points to make a meaningful extrapolation.

Let us look at the bound state mass as a
function of the coupling.
Fig.~\ref{coupling}(a) shows an example of states completely settled down
in transverse resolution, namely $K=4, T=9$. In contrast,
Fig.~\ref{coupling}(b) has only $T=1$ at $K=6$. Notice that nevertheless
both plots show extremely stable states as a function of the coupling $g$,
irrespective of their status of convergence in $T$.
As we found in our preliminary study \cite{alp99b} of this theory, all
the masses in the strong-coupling region are independent of the coupling. We
see that by a coupling of $g'=20$ a stable strong-coupling spectrum has
appeared. We have
looked as high as  couplings of 1000 and we see this same behavior for all
longitudinal and transverse resolutions.

%%%%%%%%%%%%%%%%%%%%%%%%%%
\begin{figure}[ht]
\centerline{
\psfig{file=\graphpath 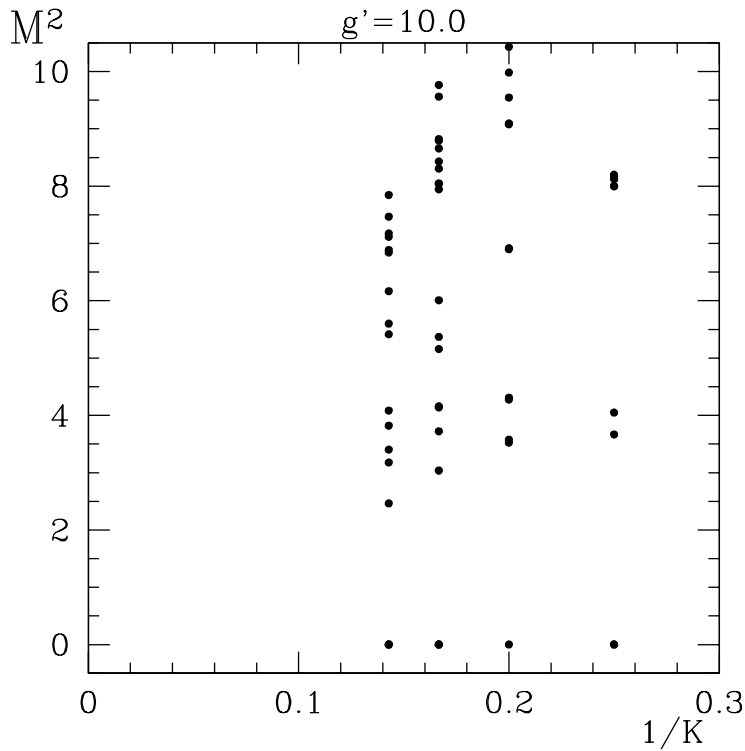,width=10.0true cm,angle=0}
}
\caption{{\small  Plot of bound state mass squared $M^2$
in units of $2 \pi^2 / L^2$  as a function of $1/K$ for coupling
$g'=10$.  For $K=3, 4$ and $5$ we plot the value
of the mass obtained by extrapolating in the transverse resolution $T$. For
$K=6$ and $7$ we
take the values at the highest resolution $T$.}
\label{long}}
\end{figure}
%%%%%%%%%%%%%%%%%%%%%%%%%%

In Figure \ref{long}  we plot the bound state mass as a function of $1/K$ .
These results are the first calculation of the
strong-coupling  bound states of ${\cal N}=1$
SYM in $2+1$ dimensions. As we increase
the resolution we are able to see states that have, 
as their primary component,
more and more partons, and, as we have seen in other supersymmetric
theories, many
of these states appear at low energies.  This accumulation of
high-multiplicity
low-mass states
appears to be a unique property of SUSY theories. In non-SUSY theories the
new states appear at increasing  energies. In the dimensionally reduced
version of this theory we saw that the
accumulation point of these low-mass states
appeared to be at zero mass
\cite{alp99b,ahlp99}. Here again we see clear evidence of an
accumulation of low
mass states, however we don't have sufficient information to say whether an
accumulation point exists.

%%%%%%%%%%%%%%%%%%%%%%%%%%
\begin{figure}[ht]
\centerline{
\psfig{file=\graphpath 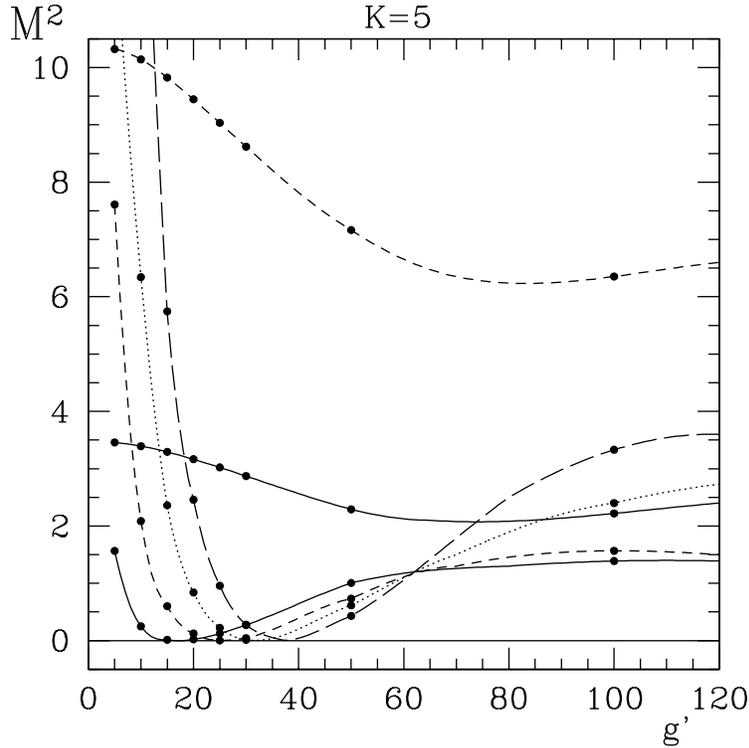,width=10.0true cm,angle=0}
}
\caption{{\small  Plot of bound state masses falling rapidly with
increasing coupling. The mass squared $M^2$ of these states
in units of $4 \pi^2/ L^2$  are plotted as a function of $g' = g \sqrt{N L}/4
\pi^{3/2}$ for $K=5$ at the transverse resolutions $T=1$ (solid lines),
$T=2$ (dashed lines), $T=3$ (short dashed line) and 
$T=4$ (long dashed line)}.}
\label{falling}
\end{figure}
%%%%%%%%%%%%%%%%%%%%%%%%%%

At this point we have not attempted to follow a specific bound state as a
function of $K$. The appearance of new low mass as we increase $K$ makes this
difficult, however a careful study of the wavefunction should make it
possible. Also the accuracy of the extrapolation to infinite $K$ would be
greatly impoved by addition of one or two additional transverse resolution at
the largest values of $K$.

\subsection{Strong Coupling: Unphysical States}

In our preliminary study of this model we found one state that was particularly
striking. It was very heavy at small coupling and approached zero mass at
strong coupling, {\em cf.}~Figure \ref{old}.  We have now been
able to look at this state at higher values of both transverse and
longitudinal resolutions. We now believe that this state is
most likely not a physical one.

We are now able to identify two states with irregular coupling dependence
which  are shown in Figure
\ref{falling} at various transverse resolution for $K=5$. We see that as we
increase the transverse cutoff these
states move up rapidly  in mass, leading us to conclude that as $T \rightarrow
\infty$, they decouple. One of these irregular states falls down to
$M=0$ and then
moves up, while another one has the same type of behavior, but with finite
minimal mass. The fact that the first state touches zero is of some interest: a
similar behavior was noted in a supersymmetric
scalar matrix model in \cite{klebhash} and in that model the authors
conjecture that this behavior might signal the existence of a critical
coupling. Here the point where the mass
goes to zero would appear to move to infinity as we remove the transverse
cutoff, and it is not at all clear that those ideas carry over here.

At $K=6$ and $K=4$ we do not see these states, at least not at the same masses
but at $K=7$ we again see such states. We believe that this is strong evidence
that these are unphysical states. Clearly a real normalizable bound state must
be visible at both even and odd resolutions, which is not the case here.
Recently we have seen unphysical massless states of this type in a two
dimensional theory with  $(8,8)$ supersymmetry \cite{ahlp99}.
In that theory we had an independent theoretical
evidence for them to be unphysical.

%%%%%%%%%%%%%%%%%%%%%%%%%%%%%%%%%%%%%%%%%%%%%%%%%%%%
\subsection{ Massless States}
In Figure \ref{old} we see a number of states that become massless
as $g' \rightarrow0$. We already explained this property in \cite{alp99b}.
Namely at zero coupling only the first term survives in the supercharge
(\ref{Qminus}) and then all the partons with $n_\perp=0$ (anti-)commute with
$Q^-$. Thus any state constructed from such partons only becomes massless.
The inverse statement is also true: at $g'=0$ a massless state cannot contain
any parton with $n_\perp\ne 0$. Thus the set of massless states at $g'=0$
coincides with a Hilbert space of the theory dimensionally reduced to $1+1$.
Moreover, the whole infrared spectrum of SYM$_{2+1}$ at small coupling is
governed by the dimensionally reduced theory (see \cite{alp99b} for details).

Previously \cite{alp99b,alp98b} we commented on the existence of
exactly massless states and in \cite{alp99b} on the one-to-one correspondence
between them and massless states of the $(1+1)$-dimensional model. Actually
this fact provides an easy way to construct massless states for
three-dimensional theories; the matrices to be
diagonalized have a size much smaller
than the ones used in the straightforward approach. The counting of
massless states
in three dimensions is also reduced to the analogous problem in $1+1$
dimensions.
For finite $N_c$, even the $(1+1)$-dimensional case is not easily handled
\cite{alp98b}; however, for large $N_c$ the multi-trace states decouple, and
%
%However even the latter is not the easy one if finite $N_c$ is considered
%\cite{alp98b}. At the large $N_c$ however the multitrace states decouple and
%
one needs to count only single-trace massless states. At resolution $K$ there
are $2(K-1)$ of them. As a numerical check of the correspondence between
massless sectors of $(2+1)$ and $(1+1)$-dimensional
theories, we can count the massless states
at different values of transverse resolution and as anticipated we found this
number to be independent of $T$.
These massless states are BPS states in the
sense that they are destroyed by one super-charge, $Q^-$, and not the
other, $Q^+$, and
the BPS bound, which is zero here, is saturated.

In \cite{alp99b} it appeared that there were additional states that became
massless as $g \rightarrow \infty$ but now we believe that these states are
unphysical. Therefore the only massless states at nonzero coupling are the
same $2(K-1)$ BPS states we saw in the dimensionally reduced model.

%%%%%%%%%%%%%%%%%%%%%%%%%%%%%%%%%%%%%%%%%%%%%%%%%%%%%%
\section{Discussion}
\label{summary}
In this work, we considered the bound states of three dimensional  SU($N$)
${\cal N}=1$ super-Yang-Mills defined on the compactified space-time ${\bf R}
\times S^1 \times S^1$. In particular, we compactified the light-cone
coordinate
$x^-$ on a light-like circle via DLCQ, and wrapped the remaining  transverse
coordinate
$x^{\perp}$ on a spatial circle. We showed explicitly that SDLCQ,  employed
in recent studies of $(1+1)$-dimensional supersymmetric gauge theories, extends
naturally to $2+1$ dimensions. The supersymmetry becomes
${\cal N}=(1,1)$ because we can always chose $P_\perp$ to be zero in  a
light-cone quantized field theory, and SDLCQ provides a regularization
scheme  that preserves this supersymmetry. The supersymmetric theory considered
here is finite  and requires no renormalization.

By retaining a finite number of transverse and longitudinal modes, we were
able to solve for bound-state wave functions and masses numerically by
diagonalizing
the discretized light-cone supercharge.  The theory clearly has a stable
spectrum at both small and large couplings. In Figure \ref{perpmom} we
see that the theory converges very rapidly in the transverse
resolution. We have
seen in reference \cite{alp98b} that SDLCQ  gives very smooth behavior in the
longitudinal resolution. In Figure \ref{falling} we see that the states with
irregular coupling dependence move off rapidly to high mass with increasing
transverse resolution. They also do not seem to appear at resolutions 4 and 6.
We therefore conclude that they are not physical states.
We see that there are no new massless states at strong coupling,
and the complete massless sector of SYM$_{2+1}$ is determined by the
two-dimensional model. The number of exactly massless states at any coupling is
$2(K-1)$, with no dependence on the transverse resolution. In addition, some
states become massless as $g$ goes to zero, but their behavior is
also described
by the theory in
$1+1$ dimensions. Consequently, we conclude that
the entire massless spectrum of the $(2+1)$-dimensional model is determined by
the dimensionally reduced model.

In previous work
\cite{alp99b} we saw that the average number of particles
in the massless states increases with
$g$ and quickly becomes equal to the maximum number allowed by the resolution.
We also see  here that the number of low-mass states increases with resolution.
Together this implies that at strong coupling the light states of this theory,
and other SUSY theories, have a huge number of degrees of freedom. No doubt,
it is this fact that allows for the possibility that
these SUSY theories can contain all of the physics of dual theories in a
different number of space-time dimensions.
It would be interesting to relate this observation with the recent
claim that strongly coupled super-Yang-Mills theory corresponds to string
theory in an anti-de Sitter background \cite{maldacena}. Of course, the
techniques we have employed here may be applied to any supersymmetric gauge
theory defined on a suitably compactified space-time. This should facilitate a
more general study of the strongly coupled dynamics of super-Yang-Mills
theories, and in particular, allow us to scrutinize the string-like properties
of Yang-Mills theories.

We have neglected the zero modes totally in this calculation. It remains an
important problem to include these modes. We already know a
great deal about zero modes \cite{alpt98} which are, after all, really only
a few extra quantum mechanical degrees of freedom. It has recently been
conjectured by Witten \cite{witten} that including these zero modes leads to
spontaneous supersymmetry breaking in this theory.

Let us briefly comment on the  $L$ dependence of the spectrum. For large
$L$ the appropriate dimensionless $L$ independent mass scale is;
\[
{M^2 \over g^4 N^2} \propto {M^2 L^2/4\pi^2 \over(g')^4}
\]
At large $g'$ the spectrum is constant in terms of $M^2
L^2/4\pi^2$, therefore in terms of this dimensionless mass scale the entire
spectrum would go
to zero as $g' \rightarrow \infty$. A possible interpretation of this result is
that in the continuum limit this theory approaches a conformal field
theory. There are however very massive states in our spectrum, that we did not
study, that could remain massive in the contimuum limit.

    The code that we are currently using is a newly
written C++ version  of the Mathematica code that we used in much of
our earlier
work. Our large runs  for matrix generation are currently performed
at the Ohio Supercomputer Center.  Matrix diagonalization is
done using standard Lapack routines \cite{Lapack} on supercomputers at the Ohio
Supercomputer Center and the Minnesota Supercomputing Institute. This is
our first  project with this new configuration, and we are currently working
on several  analytical and numerical improvements that we expect will
allow us to increase by several factors of 10 the size of the problems we
will be able to address in the future. Among these are $N=4$ SYM in $3+1$ and
$N=1$ SYM in $2+1$ with a Chern-Simons term \cite{witten}.

\medskip
{\large \bf Acknowledgments}
This work is supported in part by the US Department of Energy. We would like to
acknowledge the Ohio Supercomputer Center and the Minnesota Supercomputing
Institute for grants of computing time. We also would like to
acknowledge David G.~Robertson of the Ohio Super Computing Center
for his valued assistance and Aki Hashimoto for valuable conversations.

%%%%%%%%%%%%%%%%%%%%%%%%%%%%%%%%%%%%%%%%%%%%%%%%%%%%%%%%%%
%%%%%%%%%%%%%%%%%%%%%%%%%%%%%%%%%%%%%%%%%%%%%%%%%%%%%%%%%%

%%%%%%%%%%%%%%%%%%%%%%%%%%%%%%%%%%%%%%%%%%%%%%%%%%%%%%%

\end{document}